\documentclass[reprint,
superscriptaddress,
longbibliography,
amsmath,amssymb,
aps,
prl,
]{revtex4-1}

\usepackage{graphicx}
\usepackage{dcolumn}
\usepackage{bm}

\usepackage[colorlinks,linkcolor=blue,citecolor=blue,urlcolor=blue]{hyperref}

\usepackage{textcomp} 

\usepackage{gensymb}

\usepackage{color}

\begin{document}

\title{Measurement of the Photon-Plasmon Coupling Phase}

\author{Akbar Safari}
\email[]{asafa055@uottawa.ca}
\affiliation{Department of Physics, University of Ottawa, Ottawa, ON, K1N 6N5, Canada.}

\author{Robert Fickler}
\affiliation{Department of Physics, University of Ottawa, Ottawa, ON, K1N 6N5, Canada.}
\affiliation{Institute for Quantum Optics and Quantum Information (IQOQI), Austrian Academy of Sciences, Boltzmanngasse 3, A-1090 Vienna, Austria}

\author{Enno Giese}
\affiliation{Department of Physics, University of Ottawa, Ottawa, ON, K1N 6N5, Canada.}
\altaffiliation[]{Current affiliation: Institut f{\"u}r Quantenphysik and Center for Integrated Quantum Science and Technology (IQ$^\mathrm{ST}$), Universit{\"a}t Ulm, Albert-Einstein-Allee 11, D-89069 Ulm, Germany}

\author{Omar S. Maga\~{n}a-Loaiza}
\affiliation{Department of Physics and Astronomy, Louisiana State University, Baton Rouge, LA 70803, USA.}

\author{Robert W. Boyd}
\affiliation{Department of Physics, University of Ottawa, Ottawa, ON, K1N 6N5, Canada.}
\affiliation{Institute of Optics, University of Rochester, Rochester, New York, 14627, USA.}
\affiliation{School of Physics and Astronomy, University of Glasgow, Glasgow G12 8QQ, UK.}

\author{Israel De Leon}
\affiliation{School of Engineering and Sciences, Tecnol\'{o}gico de Monterrey, Monterrey, Nuevo Leon 64849, Mexico}

\vspace{3mm}

\date{\today}

\begin{abstract}
Scattering processes have played a crucial role in the development of quantum theory. In the field of optics, scattering phase shifts have been utilized to unveil interesting forms of light-matter interactions. Here, we investigate the mode-coupling phase of single photons to surface plasmon polaritons in a quantum plasmonic tritter. We observe that the coupling process induces a phase jump that occurs when photons scatter into surface plasmons and vice versa. This interesting coupling phase dynamics is of particular relevance for quantum plasmonic experiments. Furthermore, it is demonstrated that this photon-plasmon interaction can be modeled through a quantum-mechanical tritter. We show that the visibility of a double-slit and a triple-slit interference patterns are convenient observables to characterize the interaction at a slit and determine the coupling phase. Our accurate and simple model of the interaction, validated by simulations and experiments, has important implications not only for quantum plasmonic interference effects, but is also advantageous to classical applications. 
\end{abstract}
\maketitle
Light can couple to collective charge oscillations at the interface between a metal and a dielectric, forming surface electromagnetic waves that propagate along the interface~\cite{Raether88}. Such surface waves, referred to as surface plasmon-polaritons (SPPs), exhibit remarkable properties that make them suitable for a variety of applications~\cite{Barnes,Schuller,DeLeon14,Homola03,Lal07}. Since SPPs show intriguing non-classical effects, there is growing interest in the application of SPPs in quantum systems~\cite{Tame}. Since SPPs preserve both entanglement and photon number statistics~\cite{Lukin, Kolesov, Altewischer}, they constitute an alternative for on-chip quantum circuitry. Although SPPs are formed from photons (bosons) and electrons (fermions), they exhibit a bosonic behavior in the limit of many electrons~\cite{Pines}. Therefore, two indistinguishable SPPs interfering at a plasmonic beam splitter can bunch and show the Hong-Ou-Mandel (HOM) effect~\cite{Dheure, Martino}. In contrast to their all-optical counterpart, plasmonic beam splitters are lossy. However, these intrinsic losses are beneficial for controlling dissipative quantum dynamics, and for providing a new degree of freedom. By incorporating this additional degree of freedom, the phase shift imprinted by a plasmonic beam splitter can be adjusted such that the two SPPs antibunch~\cite{Vest}, in contrast to the conventional HOM bunching.

Similar to a scattering process, the electromagnetic field experiences a phase jump during coupling to SPPs. Determining this coupling phase and characterizing the complex photon-plasmon coupling amplitude is of great importance in designing experiments that contain quantum features such as the HOM effect. In fact, this coupling phase, also known as the scattering phase shift, is intrinsic to any scattering phenomena; a wavepacket scattering off a potential acquires a phase shift, and consequently a time delay known as the Wigner delay~\cite{Wigner,Lagendijk,Bourgain}. In plasmonic systems, this phase has been measured by employing some special techniques to image SPPs directly~\cite{Seideman, Lemke} or using several double-slit structures with different slit separations~\cite{Lezec}. However, there are inconsistencies between experimental, numerical and theoretical predictions~\cite{Hooft, Lezec, Weiner2007, Weiner2008, Seideman, Bauer}.

In this Letter, we show that the unique interference pattern of a plasmonic triple-slit is a convenient observable from which the photon-plasmon coupling phase jump can be inferred. We use a combination of double- and triple-slit structures to characterize the complex coupling amplitude. As for the quantum mechanical description of the structure, we demonstrate that each slit on a plasmonic layer can be modeled by a tritter, i.e. a device that couples three input to three output modes. Finally, we verify the accuracy of our analysis by performing a numerical simulation. Such a simple and accurate model is beneficial for future quantum plasmonic experiments.

Multiple-slit experiments lie at the heart of fundamental quantum mechanics. For example, double-slit experiments play an important role in revealing and understanding the wave-particle duality~\cite{Feynman}. Triple-slit interference patterns have been used to test the validity of Born's rule~\cite{Sinha, Sinha2, Raedt,Rahul}, one of the foundations of quantum physics. A triple-slit structure on a metallic film reveals that an additional coupling of the slits through the SPP modes leads to exotic trajectories of the pointing vector through the slit configuration and modifies the pattern, still in agreement with Born's rule~\cite{Omar}. Moreover, plasmonic slits are used to perform weak measurements~\cite{Gorodetski} and to control the spatial coherence of light~\cite{Gbur1, Gbur2, Divitt, Domenico}. Therefore, having an accurate and simple model for the coupling process at plasmonic slits can have important fundamental and practical implications.

\begin{figure}[t]                    
\begin{center}
\includegraphics[width=8.6cm]{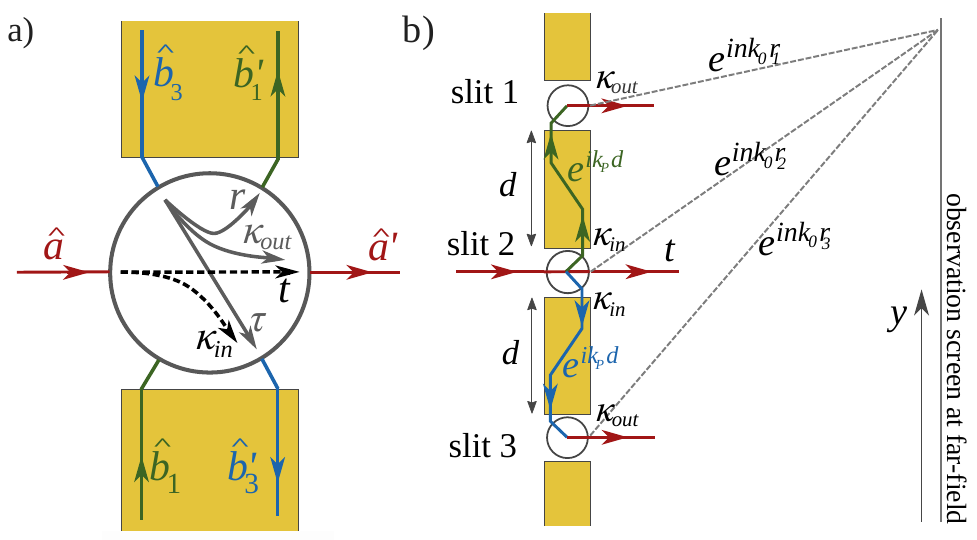}
\end{center}
\caption{Illustration of plasmonic tritters. a) Sketch of a slit on a gold film that acts as a tritter. The input field couples into two plasmonic modes each with a complex probability amplitude $\kappa_{in}$. A plasmonic mode propagating towards a slit can either reflect back, tunnel through the slit, or scatter into photons with probability amplitudes $r$, $\tau$, and $\kappa_{out}$, respectively. For clarity, we only show the coupling of one plasmonic mode. The other mode couples in a similar manner. b) Schematic diagram of the triple-slit structure where each slit acts as a tritter. The middle-slit is illuminated with single photons. The SPPs propagating from the middle-slit towards the outer slits acquire a factor of $e^{ik_Pd}$. The distance from slit $j$ to the screen at the far-field is shown by $r_j$.}
\label{Fig1} 
\end{figure} 

In contrast to the two-mode coupling typically observed in recent quantum plasmonics experiments~\cite{Dheure, Martino}, illuminating a single slit in a metallic surface can lead to a three-mode interaction, where the light field couples to two SPP modes, each of them on either side of the slit, or is transmitted through the slit, see Fig.~\ref{Fig1}\,(a). For a quantum description, we model each slit by a six-port element, a \emph{tritter}~\cite{Weihs96,Zukowski97,Schnabel06}, as a generalization of and in analogy to a beam splitter. Such elements play a crucial role for many-particle and high-dimensional quantum communication and computation~\cite{Menssen17,Nicolo,Schaeff}. In most implementations a tritter is composed of beam splitters within a complex setup~\cite{Zukowski97} or custom-tailored with integrated waveguide structures~\cite{Meany,Menssen17,Nicolo,Schaeff}, whereas the three-mode interaction at a plasmonic slit happens quite naturally.  

We introduce the six-port coupling matrix of a tritter to model the plasmonic slit and denote the input modes through the annihilation operators $\hat{a}$, $\hat{b}_1$, and $\hat{b}_3$ of the light field as well as two SPPs, respectively. We require the annihilation operators to fulfill the bosonic commutation relations. These operators are connected to the respective output modes $\hat{a}'$, $\hat{b}_1'$, and $\hat{b}_3'$ through the transformation
\begin{equation}                 
\label{e_coupling}
\begin{pmatrix}
\hat{b}_1'\\
\hat{a}'\\
\hat{b}_3'
\end{pmatrix}
=
\begin{pmatrix}
\tau	& \kappa_\text{in}	&	r	\\
\kappa_\text{out}  &	t	&	\kappa_\text{out}	\\
r	& \kappa_\text{in}	& \tau
\end{pmatrix}
\begin{pmatrix}
\hat{b}_1\\
\hat{a}\\
\hat{b}_3
\end{pmatrix}
\end{equation}
in the Heisenberg picture, see Fig.~\ref{Fig1}\,(a). Note that even though the elements of this matrix may be complex, the matrix itself has to be unitary to preserve the bosonic commutation relation and by that to conserve energy. We have also assumed that the coupling of the photon to the two SPP modes is symmetric. We perform our study at a single-photon level to lay the basis for future experiments with plasmonic slits in the quantum regime. We emphasize that at a single-photon level and to observe quantum effects such a description is necessary. However, since we only measure first moments, the same results could be obtained by using a classical light source, i.e. a laser. 

To investigate the validity of our description and to determine some of the matrix elements of Eq.~\eqref{e_coupling}, we use a triple-slit arrangement to make use of all three output channels. In general, the interference pattern generated by a triple-slit, which is essentially generated through three-path interference, can be described by the relation
\begin{align}                   
\begin{split}
I=&I_1+I_2 + I_3+2\sqrt{I_1I_2}\cos\phi_{12} \\
&+ 2\sqrt{I_2I_3}\cos\phi_{32}+2\sqrt{I_1I_3}\cos\phi_{13},
\label{e_interference} 
\end{split}
\end{align}
where $I_j$ is the intensity of the light emerging from slit $j$ and $\phi_{ij}$ is the phase difference between path $i$ and path $j$, see Fig.~\ref{Fig1}\,(b).

\begin{figure*}[ht]                    
\begin{center}
\includegraphics[width=12.8cm]{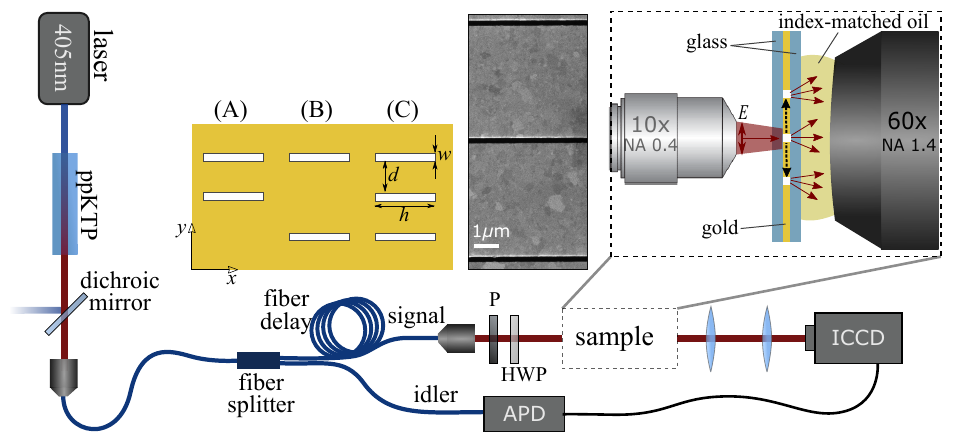} 
\end{center}
\caption{Scheme of the experimental setup. A 405-nm laser pumps a nonlinear ppKTP crystal to generate the signal-idler pairs through the process of SPDC. The idler photons are used to herald the presence of the signal photons, which are focused onto the sample by means of a microscope objective. A sketch of the sample with different slit arrangements (A), (B), and (C) is shown in the inset along with a scanning electron micrograph of the triple-slit structure of arrangement (C). The dimensions are: $w=0.20\,$\textmu m, $d=4.43\,$\textmu m, and $h=98\,$\textmu m with an uncertainty of $\pm0.03$\,\textmu m. The polarization of the signal photons is controlled by a polarizer (P) and a half-waveplate (HWP). Photons at the far field are collected with an oil-immersion objective. A lens system images the far field pattern onto an ICCD camera.}
\label{Fig2} 
\end{figure*} 

We focus on the case where only the middle slit (slit 2) is illuminated by single photons. An interference pattern forms at the far-field which can be understood by the following analysis: The photons are either coupled to two SPP modes or are transmitted through the slit. The transmission probability is $|t|^2 = I_2$ and corresponds to the normalized transmitted intensity of slit 2. The probability to couple to each plasmonic mode is $|\kappa_\text{in}|^2$. During the coupling process, the generated SPPs pick up a phase $\phi_\text{in} = \arg \kappa_\text{in}$. The losses inside the plasmonic material could be modeled by a beam splitter transformation that couples to a vacuum. However, since we are interested only in first-order moments, it is sufficient to multiply each SPP state with a factor $e^{ik_{P}d}$ to describe the propagation between the slits. Here, $d$ is the shortest distance between the outer slits and the center slit, and $k_P=k_P'+ik_P''$ is the complex wavenumber of the SPPs. When the SPPs reach the outer two slits, they can be scattered into a photonic mode with a probability of $|\kappa_{out}|^2$ and pick up a phase $\phi_\text{out}=\arg \kappa_\text{out}$. Hence, we find $I_{1}=I_3 = |\kappa_\text{in}|^2\exp(- 2 k_P'' d) |\kappa_\text{out}|^2$ for the intensity output of slits 1 and 3. From each slit the photons propagate to the screen, which gives an additional phase factor of $\exp(i n k_0 r_j )$, where $r_j$ is the distance from slit $j$ to the observation point on the screen and $n=1.52$ is the refractive index of the index-matching oil and the glass used in our microscope setup. Hence, the phase differences in Eq.~\eqref{e_interference} are given by
\begin{eqnarray}                      
\phi_{j2}=k_P'd + \phi_\text{in} + \phi_\text{out} + nk_0(r_j-r_2)
\label{eq:phase12} 
\end{eqnarray}
for $j = 1,3$ and
\begin{eqnarray}                         
\phi_{13}=\phi_{12}-\phi_{32}.
\label{eq:phase13} 
\end{eqnarray}
We demonstrate in the following that we can extract the contribution $\phi_{in}+\phi_{out}$ from the visibility of a triple-slit interference pattern.

As shown in Fig.~\ref{Fig2}, our sample contains two different double-slit structures (A) and (B) with a slit separation of $d=4.43\,$\textmu m and $9.05 \,$\textmu m, respectively, and a triple-slit structure (C) with a slit separation of $d=4.43\,$\textmu m. The sample is made of a 110-nm-thick gold film deposited on a glass substrate whose thickness is $\sim$170\,\textmu m. The complex wavenumber of the SPPs is given by~\cite{Raether88}
\begin{eqnarray}                     
k_P=k_P'+ik_P''=k_0 \sqrt{\frac{\epsilon_d \epsilon_m}{\epsilon_d+\epsilon_m}},
\label{eq:kP} 
\end{eqnarray}
where $k_0$ is the photon wavenumber in vacuum, and $\epsilon_d$ and $\epsilon_m$ are the complex relative dielectric constants of the dielectric and metal, respectively. These values are tabulated in Palik's compendium~\cite{Palik} from which we obtain $k_P=1.22 \times 10^7 +  3.39 \times 10^4 i$ for gold-glass interface at 810 nm as used in our experiment. Note that the film is thick enough to avoid coupling between the SPP modes excited on the top and bottom surfaces of the film~\cite{Raether88}. In our experiments, we illuminate one of the slits with heralded single photons focused by a microscope objective on the sample. An index-matching oil-immersion microscope objective is utilized to magnify the field distribution and an imaging system images the far field pattern onto an intensified charge-coupled device (ICCD) camera. 

Our heralded single-photon source is realized using spontaneous parametric down-conversion (SPDC) in a 2-mm-long type-I periodically poled potassium titanyl phosphate (ppKTP) nonlinear crystal pumped by a 405\,nm continuous wave diode laser ($\sim$200\,mW). The pairs are degenerate at a wavelength of 810\,nm and pass through a 3-nm-band-pass filter before they couple into a single-mode fiber. The idler and the signal photons are separated probabilistically by means of a 50/50 fiber beam splitter. A coincidence count rate of $\sim$36\,kHz is obtained. We detect the idler photons with a single-photon avalanche photo diode (APD) that is used to trigger the ICCD camera that registers the detection of the signal photons. To compensate for the electronic delay of the camera, we delay the signal photons by passing them through a 22-m-long fiber before we send them through the sample. The ICCD (with a 7-ns-gate-time) registers the signal photons in the far field of the slits. 

\begin{figure}[t]                    
\begin{center}
\includegraphics[width=8.6cm]{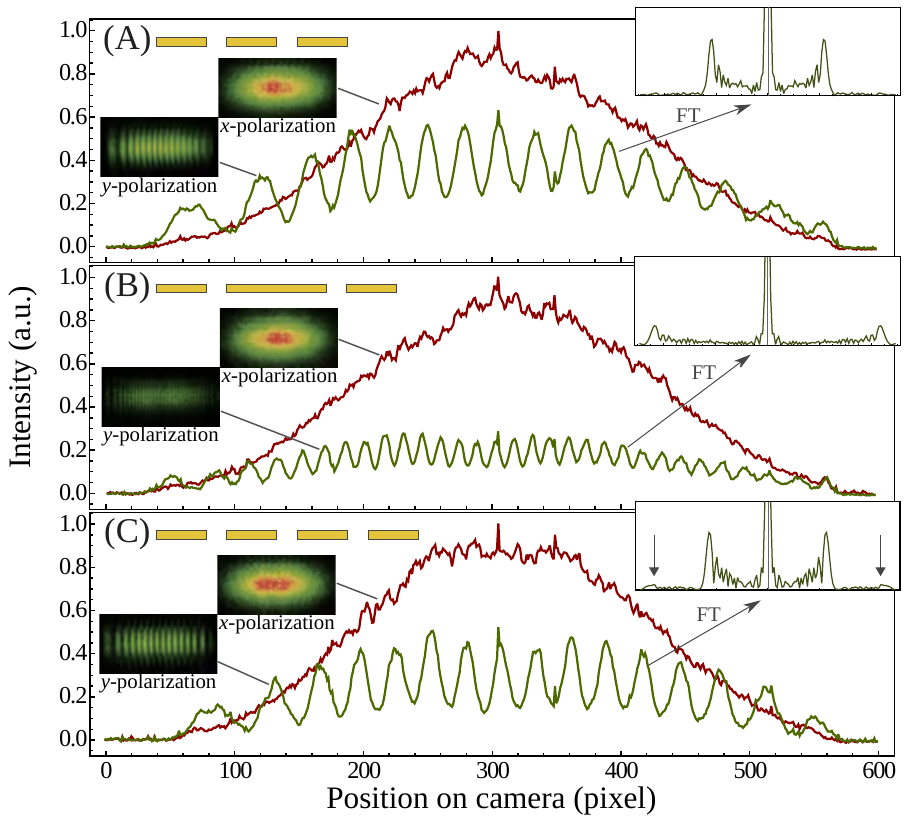} 
\end{center}
\caption{Far-field interference patterns from the three different slit structures labeled in Fig.~\ref{Fig2}. Only the photons with $y$-polarization excite SPPs (green), no interference occurs for $x$-polarization (red). The Fourier transforms (FT) in each part show the spatial frequency of the fringes. Since the slit separation is larger in (B) its fringe pattern has a higher spatial frequency than (A). In (C), the interference of the two modes emerging from the two outer slits has a small contribution in the triple-slit pattern as shown with the arrows on the Fourier transform.}
\label{Fig3} 
\end{figure}

The excitation of SPPs at a slit requires a transverse magnetic polarization. If the photons are polarized along the long axis of the slit ($x$-polarization), there is no coupling to plasmonic modes and the far-field pattern does not show any interference; see the red patterns in Fig.~\ref{Fig3}. However, upon rotation of the polarization of the input photons by 90 degrees an interference pattern is formed even though only one slit is illuminated~\cite{Ravets,Mori,Wang}. Multiple-slit interference occurs because the SPPs excited at the illuminated slit propagate to the neighboring slits where they scatter into photons. The measurements depicted in green in Fig.~\ref{Fig3} show the interference pattern for the polarization perpendicular to the long axis of the slit ($y$-polarization). For the rest of the experiment we perform the measurements with $y$-polarized photons to excite SPPs and use the $x$-polarized photons only to calibrate the far field pattern.

To obtain the modulus of the photon-plasmon coupling constant of Eq.~\eqref{e_coupling}, we first measure the visibility of the double-slit structure (A). We analyze our data with Eq.~\eqref{e_interference} and set $I_3=0$. Equation~\eqref{e_interference} therefore reduces to a simple double-slit pattern $I = I_1+I_2 + 2\sqrt{I_1I_2} \cos \phi_{12}$ with a visibility of $V = 2 \sqrt{I_1 I_2}/(I_1+I_2)$. Since $I_2=|t|^2$ and $I_1 = |\kappa_\text{in}|^2\exp(- 2 k_P'' d) |\kappa_\text{out}|^2$ with $d= 4.43\,$\textmu m, the visibility depends on the three coupling parameters $|t|$, $|\kappa_{in}|$, and $|\kappa_{out}|$. However, because of the unitarity of the tritter matrix we have $|t|^2 = 1 - 2 |\kappa_\text{in}|^2$. If we additionally assume reciprocity of the coupling process~\cite{Schnabel06}, we find $|\kappa_\text{in}|=|\kappa_\text{out}|$ and the visibility depends only on one free parameter.

We measure a visibility $V=0.41\pm0.01$ for slit structure (A) from the interference in Fig.~\ref{Fig3}(A) and extract $|t|=0.78\pm0.01$ and $|\kappa_{in}|=|\kappa_{out}|=0.44\pm0.01$. For a consistency check, we measure the visibility of the double-slit structure (B) from the pattern shown in Fig.~\ref{Fig3}(B) and we obtain $V=0.35\pm0.01$.
The theoretical prediction based on the values determined above and with $d=9.05$\,\textmu m is $V=0.34\pm0.01$, which shows a prefect agreement to our experimental result.

\begin{figure}[t]                    
\begin{center}
\includegraphics[width=8.6cm]{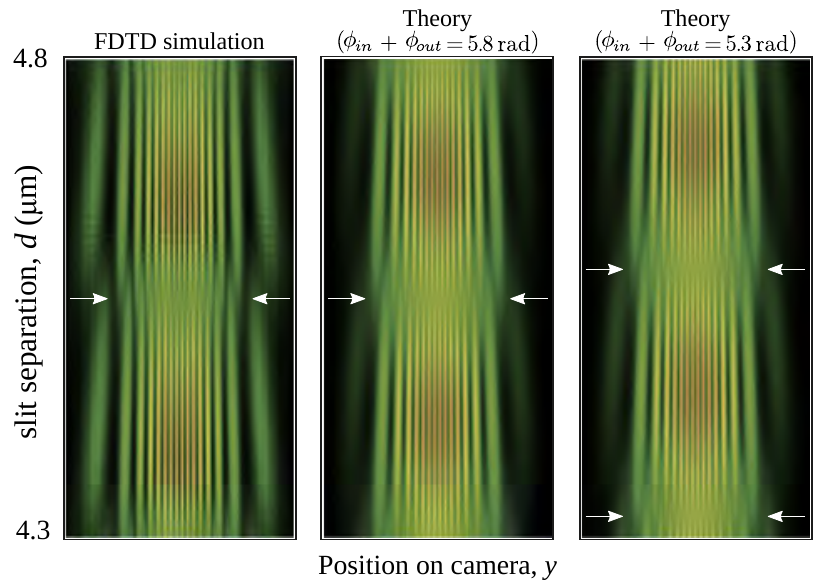} 
\end{center}
\caption{Far-field interference pattern for different slit separations from $4.3$\,\textmu m to $4.8$\,\textmu m obtained from FDTD simulation and from our theoretical model. For the theoretical plots Eq.~\eqref{e_interference} is multiplied by a sinc function to account for the finite width of the slits. The theoretical result matches to the numerical simulation when we incorporate a coupling phase of $\phi_{in}+\phi_{out}=5.8$ radians. With other values of the coupling phase the position of the minimum visibility (indicated by the arrows) shifts, as shown in the rightmost plot. The excellent agreement between the simulation and the theoretical results confirms the validity of our theoretical model. The outer fringes in the theoretical plots are faint due to the deviation from the small angle approximation used for the sinc function.}
\label{Fig4} 
\end{figure} 

In a double-slit experiment, the visibility is independent of the coupling phases $\phi_\text{in}$ and $\phi_\text{out}$, since they only appear as a transverse shift of the far-field interference pattern. Finding the zero-fringe to obtain these coupling phases in a double-slit experiment would require a perfect alignment of the camera to the center of the structure. To avoid this difficulty and use only the visibility as an observable, we turn to a triple-slit structure (C) to measure the coupling phases. The visibility of the triple-slit pattern depends on the coupling phases because the far-field intensity has three interference contributions, as demonstrated by Eq.~\eqref{e_interference}. We assume that the device is symmetric so that $I_1=I_3 < I_2$, and, through use of a trigonometric identity, Eq.~\eqref{e_interference} reduces to
\begin{equation}                     
\label{e_int_triple}
I = 2 I_1+ I_2 + 4 \sqrt{I_1 I_2} \cos\frac{\phi_{12}+\phi_{32}}{2} \cos \frac{\phi_{13}}{2} + 2 I_1 \cos \phi_{13},
\end{equation}
where we used the definition of $\phi_{13}$ from Eq.~\eqref{eq:phase13}. Since $\phi_{12}+\phi_{32}$ depends only weakly on the transverse position in the far-field, the first cosine in Eq.~\eqref{e_int_triple} does not vary significantly over the interference pattern. Also, the last term in Eq.~\eqref{e_int_triple} makes a small contribution as $I_1 \ll 2 \sqrt{I_1 I_2}$. This conclusion can be drawn by looking at the Fourier transform of the experimental interference pattern shown in the inset of Fig.~\ref{Fig3}\,(C); the dominant spatial frequency comes from $\cos(\phi_{13}/2)$, and the contribution of $\cos\phi_{13}$, which oscillates at twice this frequency, is negligible. Hence, the visibility is determined by $\phi_{12}+\phi_{32}$, which includes the coupling phases as well as the phase $k'_P d$.

By matching the visibility of our model to the visibility of the experimental pattern shown in Fig.~\ref{Fig3}\,(C), we find the coupling phases to be $\phi_\text{in}+\phi_\text{out}=5.4\pm0.4$\,radians. Note that the main source of uncertainty comes from the measurement of the slit separation from the scanning electron micrograph in Fig.~\ref{Fig2}. 

To test the accuracy of our theoretical model and its results, we perform a finite-difference time-domain (FDTD) simulation of the triple-slit experiment. Since the visibility of the interference pattern depends on the exact size of the beam on the illuminated slit and absolute coupling efficiencies, the visibility of the simulated pattern cannot be compared directly with that of the experiment. Therefore, we plot the far-field pattern for different slit separations $d$, from $4.3\,$\textmu m to $4.8\,$\textmu m. By comparing the results of the FDTD simulation and our theoretical model shown in Fig.~\ref{Fig4}, we find that the coupling phase is $\phi_\text{in}+\phi_\text{out}=5.8\pm0.1$\,radians. This value is in complete agreement with the coupling phase obtained from the experimental data. If we were to use a different coupling phase in our theoretical model, the position of the minimum of visibility would be shifted. As an example, Fig.~\ref{Fig4} also shows the theoretical pattern with a wrong coupling phase of 5.3\,radians. The minimum visibility is clearly shifted upwards and another minimum appears from the bottom of the pattern. 

In summary, we have characterized three-mode photon-plasmon coupling phenomena at a slit by employing a simple but accurate quantum-mechanical description of a tritter. We showed that a triple-slit arrangement constitutes a convenient structure to analyze the six-port coupling matrix, and in particular to measure the phase of the coupling process. Using our model, we experimentally observed that photons experience a significant phase jump of $5.4\pm0.4$\,radians as they couple and de-couple to and from the plasmonic structure. This phase jump is a generic physical phenomenon and occurs in any scattering event. Thus, our approach provides an easy method to measure this scattering phase and its dependence on different parameters. As this coupling phase is of particular importance in observing HOM-like quantum interference effects, we anticipate that the results of our study will be relevant for future quantum-plasmonic experiments. Last but not least, the complex nature of the multi-mode photon-plasmon coupling at a slit might lead to the investigation of unique properties of multi-particle interactions~\cite{Menssen17}.

\subsection{Acknowledgment}
This work was supported by the Canada Excellence Research Chairs program and the National Science and Engineering Research Council of Canada (NSERC). R.F. also acknowledges the support of the Banting postdoctoral fellowship of NSERC. 

\bibliography{Ref}

\end{document}